\documentclass[aps,nofootinbib,prd,superscriptaddress,tightenlines,notitlepage,twocolumn,showpacs]{revtex4-1}

\usepackage[colorlinks=true, linkcolor=blue, citecolor=blue]{hyperref}
\usepackage{tabularx}
\usepackage{graphicx}
\usepackage{amssymb,bm}
\usepackage{xcolor}
\usepackage{amsmath}
\usepackage[varg]{txfonts}
\usepackage{enumerate}
\usepackage[normalem]{ulem}
\usepackage[all]{hypcap}
\usepackage{subcaption}

\begin{document}

%\linenumbers

\title{Stringent Tests of Lorentz Invariance Violation from LHAASO Observations of GRB 221009A}

%%%%%%%%%%%%%%%%%%%%%%%%%%%%%%%%%%%%%%%%%%%%%%%%%%%%%%%%%%%author list start

\author{Zhen Cao}
\affiliation{Key Laboratory of Particle Astrophysics \& Experimental Physics Division \& Computing Center, Institute of High Energy Physics, Chinese Academy of Sciences, 100049 Beijing, China}
\affiliation{University of Chinese Academy of Sciences, 100049 Beijing, China}
\affiliation{Tianfu Cosmic Ray Research Center, 610000 Chengdu, Sichuan,  China}

\author{F. Aharonian}
\affiliation{Dublin Institute for Advanced Studies, 31 Fitzwilliam Place, 2 Dublin, Ireland }
\affiliation{Max-Planck-Institut for Nuclear Physics, P.O. Box 103980, 69029  Heidelberg, Germany}

\author{Axikegu}
\affiliation{School of Physical Science and Technology \&  School of Information Science and Technology, Southwest Jiaotong University, 610031 Chengdu, Sichuan, China}

\author{Y.X. Bai}
\affiliation{Key Laboratory of Particle Astrophysics \& Experimental Physics Division \& Computing Center, Institute of High Energy Physics, Chinese Academy of Sciences, 100049 Beijing, China}
\affiliation{Tianfu Cosmic Ray Research Center, 610000 Chengdu, Sichuan,  China}

\author{Y.W. Bao}
\affiliation{School of Astronomy and Space Science, Nanjing University, 210023 Nanjing, Jiangsu, China}

\author{D. Bastieri}
\affiliation{Center for Astrophysics, Guangzhou University, 510006 Guangzhou, Guangdong, China}

\author{X.J. Bi}
\affiliation{Key Laboratory of Particle Astrophysics \& Experimental Physics Division \& Computing Center, Institute of High Energy Physics, Chinese Academy of Sciences, 100049 Beijing, China}
\affiliation{University of Chinese Academy of Sciences, 100049 Beijing, China}
\affiliation{Tianfu Cosmic Ray Research Center, 610000 Chengdu, Sichuan,  China}

\author{Y.J. Bi}
\affiliation{Key Laboratory of Particle Astrophysics \& Experimental Physics Division \& Computing Center, Institute of High Energy Physics, Chinese Academy of Sciences, 100049 Beijing, China}
\affiliation{Tianfu Cosmic Ray Research Center, 610000 Chengdu, Sichuan,  China}

\author{W. Bian}
\affiliation{Tsung-Dao Lee Institute \& School of Physics and Astronomy, Shanghai Jiao Tong University, 200240 Shanghai, China}

\author{A.V. Bukevich}
\affiliation{Institute for Nuclear Research of Russian Academy of Sciences, 117312 Moscow, Russia}

% ...
\collaboration{The LHAASO Collaboration}

%%%%%%%%%%%%%%%%%%%%%%%%%%%%%%%%%%%%%%%%%%%%%%%%%%%%%%%%%%%%%%%%%%%%%%%%%%end author list

\email[E-mail:]{gmxiang@ihep.ac.cn; jjwei@pmo.ac.cn; zhiguo.yao@ihep.ac.cn; xfwu@pmo.ac.cn}

\date{\today}
%\noaffiliation

\begin{abstract}
{\color{black}On October 9, 2022}, the Large High Altitude Air Shower Observatory (LHAASO) reported the observation of the very early
TeV afterglow of the brightest-of-all-time GRB 221009A, recording the highest photon statistics in the TeV band
{\color{black}ever obtained} from a Gamma-ray burst. We use this unique observation to place stringent constraints on an energy
dependence of the speed of light in vacuum, a manifestation of Lorentz invariance violation (LIV) predicted by
some quantum gravity (QG) theories. Our results show that the 95\% confidence level lower limits on
the QG energy scales are $E_{\mathrm{QG},1}>10$ times of the Planck energy $E_\mathrm{Pl}$ for the linear, and
$E_{\mathrm{QG},2}>6\times10^{-8}E_\mathrm{Pl}$ for the quadratic LIV effects, respectively. Our limits on the quadratic LIV
case improve previous best bounds by factors of 5--7.
\end{abstract}

%\pacs{05.65.+b, 98.70.Rz, 05.45.Tp}

%\keywords{}

\maketitle

\section{Introduction}
Lorentz invariance, the fundamental symmetry of Einstein's relativity, has withstood strict tests
over the past century~\citep{2011RvMP...83...11K}. However, deviations from Lorentz invariance at energies
approaching the Planck scale $E_\mathrm{Pl}=\sqrt{\hbar c^{5}/G}\simeq1.22\times10^{19}$ GeV are predicted
in many quantum gravity (QG) theories seeking to unify quantum theory and general relativity
{\color{black}\cite{1989PhRvD..39..683K,1997IJMPA..12..607A,1999GReGr..31.1257E,2002Natur.418...34A,2002PhRvL..88s0403M,
2002PhRvD..65j3509A,2005LRR.....8....5M,2009PhLB..679..407L,2013LRR....16....5A,2014RPPh...77f2901T,
2021FrPhy..1644300W,2022Univ....8..323H,2022PrPNP.12503948A,2023arXiv231200409A}}. Although {\color{black}any 
signals of Lorentz invariance violation} (LIV) are expected to be very tiny at attainable energies 
$\ll E_\mathrm{Pl}$, they can increase with energy and accumulate to detectable levels over large propagation 
distances. Astrophysical observations involving high-energy radiation and long distances are therefore 
suitable for performing sensitive tests of Lorentz invariance.

One of the manifestations of LIV can be characterised as energy-dependent modifications to the photon
dispersion relation in vacuum~\citep{1998Natur.393..763A}:
\begin{equation}
E^{2}\simeq p^{2}c^{2}\left[1-\sum_{n=1}^{\infty}s\left(\frac{E}{E_{{\rm QG},n}}\right)^{n}\right]\;,
\label{eq:LIVdispersion}
\end{equation}
where $E$ and $p$ are the energy and momentum of a photon, $s=\pm1$ is the ``sign'' of the LIV effect, corresponding to the ``subluminal'' or
``superluminal'' scenarios, and $E_{\mathrm{QG},n}$
denotes the hypothetical QG energy scale. Since the sum is dominated by the lowest-order term of the series
at small energies $E\ll E_{\mathrm{QG},n}$, only the first two leading terms ($n=1$ or $n=2$) are of interest
for independently LIV tests. They are usually referred to as linear and quadratic LIV corrections, respectively.
Taking into account only the leading LIV modification of order $n$, the photon group velocity is then given by
\begin{equation}
\upsilon(E)=\frac{\partial E}{\partial p}\approx c\left[1-s\frac{n+1}{2}\left(\frac{E}{E_{{\rm QG},n}}\right)^{n}\right].
\label{eq:vLIV}
\end{equation}
Because of the energy dependence of $\upsilon(E)$, two photons with different
energies (denoted by $E_{h}$ and $E_{l}$, where $E_{h}>E_{l}$) emitted simultaneously from the same source
at redshift $z$ would reach us at different times. The energy-dependent time delay due to LIV effects can be
expressed as~\citep{2008JCAP...01..031J}
\begin{equation}
%\begin{aligned}
\Delta t_{\rm LIV}=s\frac{n+1}{2}\frac{E^{n}_{h}-E^{n}_{l}}{E_{{\rm QG}, n}^{n}}
\int_{0}^{z}\frac{(1+z')^{n}}{H(z')}dz'\;,
\label{eq:tLIV}
%\end{aligned}
\end{equation}
where $H(z)=H_{0}\sqrt{\Omega_\mathrm{m}(1+z)^{3}+\Omega_{\Lambda}}$, assuming a flat $\Lambda$CDM cosmology with
Hubble constant $H_{0}=67.36~\mathrm{km~s^{-1}~Mpc^{-1}}$, matter density parameter $\Omega_\mathrm{m}=0.315$,
and vacuum energy density $\Omega_{\Lambda}=1-\Omega_\mathrm{m}$ \citep{2020A&A...641A...6P}. For convenience,
in Eq.~(\ref{eq:tLIV}) we introduce the dimensionless LIV parameters
\begin{equation}
\label{eq:eta1}
\eta_{1}=sE_{\rm Pl}/E_{{\rm QG},1}
\end{equation}
and
\begin{equation}
\label{eq:eta2}
\eta_{2}=10^{-15} \times s E_{\rm Pl}^{2}/E_{{\rm QG},2}^{2}
\end{equation}
for linear ($n=1$) and quadratic ($n=2$) modifications, respectively, to replace $E_{\mathrm{QG},1}$ and $E_{\mathrm{QG},2}$.

It is obvious from Eq.~(\ref{eq:tLIV}) that the greatest sensitivities on {\color{black}
$\eta_{n}$ (or $E_{\mathrm{QG},n}$)}
can be expected from those astrophysical sources with rapid signal variability, large distances, and high-energy
emission. As the most violent explosions occurring at cosmological distances, Gamma-ray bursts (GRBs) have been
deemed as excellent probes for searching for the LIV-induced vacuum dispersion~\citep{1998Natur.393..763A,2009Sci...323.1688A,2009Natur.462..331A,2013PhRvD..87l2001V,2019PhRvD..99h3009E}.
Indeed, the most stringent limits to date on {\color{black}$\eta_{n}$ (or $E_{\mathrm{QG},n}$)}, resulting from vacuum dispersion time-of-flight studies,
have been obtained using the {\color{black}GeV emission by} GRB~090510 observed by the Fermi-LAT.
The limits set for the subluminal (superluminal) scenario are {\color{black}$\eta_{1}<0.13$, or equivalently $E_{\mathrm{QG},1}>9.3\times10^{19}~\mathrm{GeV}$
($\eta_{1}>-0.09$, or equivalently $E_{\mathrm{QG},1}>1.3\times10^{20}~\mathrm{GeV}$) and $\eta_{2}<8.8$, or equivalently $E_{\mathrm{QG},2}>1.3\times10^{11}~\mathrm{GeV}$
($\eta_{2}>-16.8$, or equivalently $E_{\mathrm{QG},2}>9.4\times10^{10}~\mathrm{GeV}$) for linear and quadratic LIV effects}, respectively~\citep{2013PhRvD..87l2001V}.
Based on the detection of sub-TeV emission from GRB~190114C, {\color{black}MAGIC Collaboration (2020)~\citep{PhysRevLett.125.021301}} obtained competitive lower limits on the quadratic LIV energy scale, i.e., {\color{black}$\eta_{2}<37.0$, or equivalently $E_{\mathrm{QG},2}>6.3\times10^{10}~\mathrm{GeV}$ ($\eta_{2}>-48.0$, or equivalently $E_{\mathrm{QG},2}>5.6\times10^{10}~\mathrm{GeV}$)
for the subluminal (superluminal) case.}

The Large High Altitude Air Shower Observatory (LHAASO) detected more than 64,000 photons in the energy range of 0.2--7~TeV from GRB 221009A within the first 4000~s after the MeV burst trigger~\citep{2023Sci...380.1390L}. This object
is located at redshift $z=0.151$~\citep{2022GCN.32648....1D,2023arXiv230207891M}.
In this Letter, we study Lorentz-violating effects using the time-of-flight measurements of the unprecedentedly very-high-energy (VHE, $>100$~GeV) photons from GRB 221009A.

\section{LHAASO observations of GRB 221009A}
\label{sec:Observation}
{\color{black}At 13:16:59.99 UTC on 9 October 2022 (denoted as $T_0$), a long-duration GRB, numbered as GRB 221009A,
triggered the Gamma-ray Burst Monitor (GBM) onboard the Fermi satellite~\cite{2022GCN.32636....1V,2023ApJ...952L..42L}.}
The subsequent detection with Fermi-LAT made clear that it is an extraordinarily bright
burst~\citep{2022GCN.32637....1B,2022GCN.32658....1P}. The Gamma-ray emission of GRB 221009A was also detected by several
other space missions~\cite{2022GCN.32632....1D,2022GCN.32688....1K,2022GCN.32650....1U,
2022GCN.32657....1P,2022GCN.32660....1G,2022GCN.32661....1X,2022GCN.32663....1L,2022GCN.32668....1F,2023ApJ...949L...7F,2022GCN.32685....1R,2023A&A...677L...2R,2022GCN.32746....1M,2022GCN.32751....1L,2023arXiv230301203A,2022GCN.32973....1D}, and by the ground-based air shower detector LHAASO~\cite{2022GCN.32677....1H}.

LHAASO~\cite{2022ChPhC..46c0001M} is a new generation Gamma-ray and cosmic-ray observatory situated in Daocheng, China, at an elevation of $\sim4410$ meters. Due to the large area, wide field of view, and broad energy coverage, the LHAASO detectors are meticulously designed to delve into new frontiers physics, including investigations into LIV, among other scientific objectives.

At $T_0$, GRB 221009A was observed by LHAASO at a zenith angle of $28.1^{\circ}$, and remained within LHAASO's field of view for the next 6000 seconds. In the initial 4000 seconds, the Water Cherenkov Detector Array (WCDA) of LHAASO captured over 64,000 photons in the 0.2--7~TeV energy range, both the light curve and energy spectrum of VHE photons were measured~\cite{2023Sci...380.1390L}. The intrinsic light curve reveals a rise to a peak from 231 to 244 seconds after $T_0$, followed by a decay lasting 650 seconds.

The light curve of energy flux in the specified time range, as detected by LHAASO-WCDA, can be well described by a smoothly broken power-law function~\citep{2023Sci...380.1390L},
%The theoretical light curve of energy flux in the acceleration and the coasting stage of a canonical GRB afterglow process, corresponding to above-mentioned rise and decay phase detected by LHAASO-WCDA, can be well described by a smoothly broken power-law function~\citep{2023Sci...380.1390L},
\begin{align}
 \lambda(t) \propto \left[\left(\frac{t}{t_{\rm b}}\right)^{-\omega\alpha_1}+\left(\frac{t}{t_{\rm b}}\right)^{-\omega\alpha_2}\right]^{-1/\omega},
\label{eq:lc_func}
\end{align}
where all time-related variables are relative to a reference time $T^\ast=T_{0} + 226\,\mathrm{s}$. Here, $\alpha_{1}=1.82$ and $\alpha_{2}=-1.115$ denote the power-law indices before and after the break time $t_\mathrm{b}=15.37\,\mathrm{s}$, and $\omega=1.07$ represents the sharpness of the break.  The intrinsic time-resolved spectrum can be fitted with a power-law function, and the positive power-law spectral index varies with time following the expression~\cite{2023Sci...380.1390L}
\begin{align}
\label{eq:index_evolution}
\gamma(t) &= a \log(t) + b,
\end{align}
where the unit of $t$ is seconds, and $a=-0.135$ and $b=2.579$. When a time delay due to LIV is introduced, this $\lambda(t)$ will be modified to $\lambda\left[t-\Delta t_{\rm LIV}(E,\eta_n)\right]$.

The observed count rate light curve, characterizing the probability of observing a photon in a range of the number of hits $\Delta N_\mathrm{hit}$ and the arrival time $t$ from the GRB, can be converted from the energy flux light curve with
\begin{align}
f(t, \Delta N_{\rm hit}|\eta_n) &= \int_0^{+\infty} \lambda\left[t - \Delta t_{\rm LIV}(E,\eta_n)\right] \nonumber\\
&\times \zeta(t)\,E^{-\gamma(t)}\,P_{\rm EBL}(E)\, S(E,t,\Delta N_{\rm hit})\, {\rm d}E.
\label{eq:ratelc_func}
\end{align}
Here, $\zeta(t) = A/\int_{E_1}^{E_2} E^{1-\gamma(t)}\,{\rm d}E$ serves as the conversion factor from energy flux (within the energy range from $E_1 = 0.3\,\mathrm{TeV}$ to $E_{2}=5\,\mathrm{TeV}$) to flux coefficient, where $\gamma(t)$ represents the power-law index evolution as per Eq.~(\ref{eq:index_evolution}), and $A$ is a constant to be determined. The term $P_\mathrm{EBL}(E)$ denotes the survival probability of photons subject to extra-galactic background light (EBL) attenuation, {\color{black}adopting the model~\citep{2021MNRAS.507.5144S}}. Lastly, {\color{black}$S(E,t,\Delta N_\mathrm{hit})$ accounts for the effective detection area of photons at energy $E$ and time $t$, with the number of fired cells in a given segment $\Delta N_\mathrm{hit}$.}

\section{Analysis methods and results}
\label{sec:method}

We utilize two analysis methods to examine the LIV lags in the VHE Gamma-ray signals from GRB 221009A. The cross-correlation function (CCF) is employed to directly measure the time delays between different energy bands, while the maximum likelihood (ML) method is adopted to extract energy-dependent arrival time delays. These two methods are widely employed in similar LIV studies.

\subsection{Cross-correlation Function Method}

\begin{figure}
\vskip-0.4in
\centerline{\includegraphics[angle=0,width=0.45\textwidth]{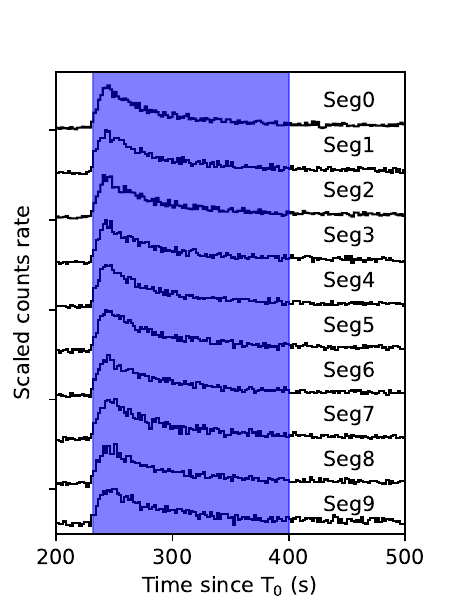}}
\caption{Count rate light curves of GRB 221009A, as detected by LHAASO-WCDA, presented in ten $N_{\rm hit}$ segments. The time binning of the light curves used for analysis is 0.1~s; however, they are depicted here with 2~s intervals for clarity. The blue-marked range 232--400~s after the GBM trigger $T_0$ is selected for calculating the time lags. }
\label{fig:ten_segments}
\end{figure}

We segment the light curve of the count rate detected by LHAASO-WDA from GRB 221009A into ten intervals, covering the time span from 232 to 400 seconds after $T_0$. The segmentation is based on the number of fired cells ($N_\mathrm{hit}$), with approximately the same number of events in each segment. The $N_\mathrm{hit}$ segments roughly correspond to different energy ranges, 
and the median energies are 0.354, 0.375, 0.395, 0.419, 0.457, 0.486, 0.556, 0.658, 0.843, and 1.601~TeV, respectively, 
considering the spectral index evolution from~\cite{2023Sci...380.1390L} for the interested time span. The energy-dependent light curves of GRB 221009A for the ten $N_\mathrm{hit}$ segments (denoted by Seg0--Seg9) are displayed in Fig.~\ref{fig:ten_segments}.

%Time lags between the lowest energy band (Seg0) and any of the other nine high energy bands (Seg1--Seg9) can be calculated using the CCF defined as
%\begin{align}
%F_j(\Delta t) &= \frac{\sum_i R_0(t_i)\,R_j(t_i+\Delta t)}{\sqrt{\sum_i R_0^2(t_i)\,\sum_i R_j^2(t_i)}},
%\label{eq:ccf}
%\end{align}
%where $j = 1, 2, 3, \ldots, 9$ represents the $N_\mathrm{hit}$ segment number except the first one, $i$ is the time bin number, $R_0(t_i)$ and $R_j(t_i)$ are the rates of segment 0 and $j$ at time bin $t_i$, and $\Delta t$ is the time lag. A discrete analysis is performed with a time step 0.1\,s. By fitting $F_j(\Delta t)$ around the peak with a Gaussian function, a more precise peak position $\Delta t_j$ can be obtained (see Section~\ref{sec:Appen1} of the Supplemental Material).

In our analysis, we use the the CCF method to calculate the time lags $\Delta t$ between the lowest energy band (Seg0) and any of the other nine high energy bands (Seg1--Seg9) (see Section~\ref{sec:Appen1} of the Supplemental Material for details).
Assuming that the observed time lags $\Delta t$ are primarily caused by LIV effects, we can establish a conservative limit on the LIV parameter $\eta_n$ ($n=1$ or $2$). Utilizing the 9 pairs of CCF measurements, we conduct a global fit to constrain $\eta_n$ by minimizing
\begin{align}
\chi^2(\eta_n) = \sum_j   \frac{  \left[\Delta t_j - \Delta t_{\rm LIV}(\eta_{n}) \right]^{2}   }{\sigma^2(\Delta t_j)}, 
\label{eq:ccf_likelihood}
\end{align}
where $\sigma(\Delta t_j)$ is the uncertainty of $\Delta t_j$, regarded as a statistical origin. This uncertainty is obtained by a bootstrapping method: mocking the data for this CCF pair 1000 times, and RMS of the obtained $\Delta t_j$ is set to $\sigma(\Delta t_j)$.

In Eq.~(\ref{eq:ccf_likelihood}), the median energy of photons is used to calculate the LIV time delay $\Delta t_\mathrm{LIV}(\eta_n)$ for each $N_\mathrm{hit}$ segment. {\color{black}However, this method may introduce bias due to several factors: spectral index evolution, wide dispersion of photon energies within each $N_\mathrm{hit}$ segment (mainly caused by air shower fluctuations), and significant overlap in energy ranges among adjacent segments (see Section~\ref{sec:Appen2} of the Supplemental Material).}

\begin{figure}
\centerline{\includegraphics[angle=0,width=0.55\textwidth]{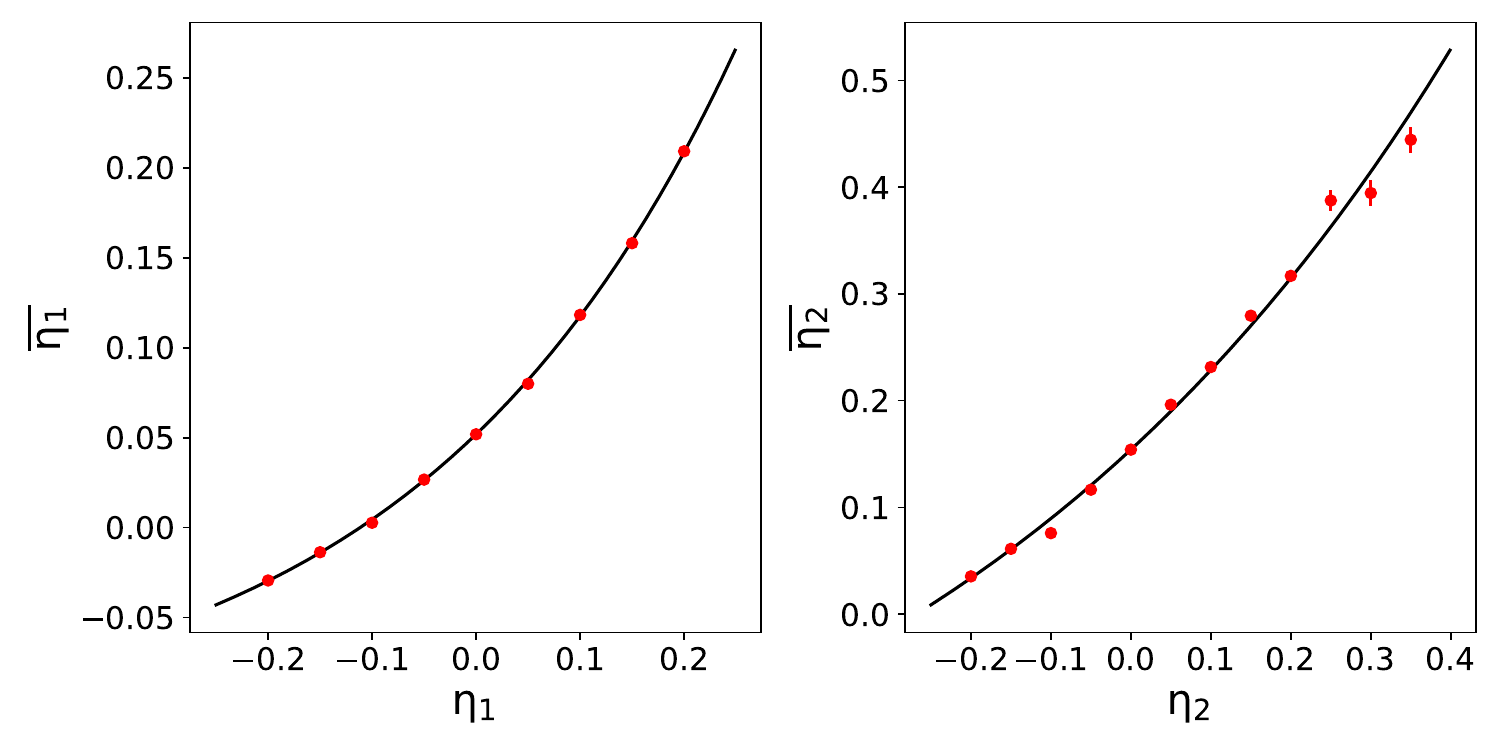}}
\caption{Relationship between the induced LIV value of $\overline{\eta_n}$ (with $\overline{\eta_1}$ in the left panel and $\overline{\eta_2}$ in the right panel) and the input $\eta_n$ (denoted as $\eta_1$ and $\eta_2$ respectively), obtained through a simulation procedure.}
\label{fig:ccf_bias}
\end{figure}

The biases are estimated as follows. Assuming a LIV parameter $\eta_1$ or $\eta_2$, the intrinsic light curve of energy flux as Eq.~(\ref{eq:lc_func}) can be simulated, and then translated by Eq.~(\ref{eq:ratelc_func}) into ten count rate light curves. Employing simulated light curves, the same procedure as with the data is executed, leading to the determination of the measured LIV parameters, as shown in Fig.~\ref{fig:ccf_bias}. This involves iterating over various $\eta_1$ and $\eta_2$ values as input, enabling a polynomial fitting. Using the fitted function, the bias $\overline{\eta_n}-\eta_n$ for the analysis results on experimental data can be evaluated and subtracted.

\begin{figure}
%\vskip-0.1in
\centerline{\includegraphics[angle=0,width=0.5\textwidth]{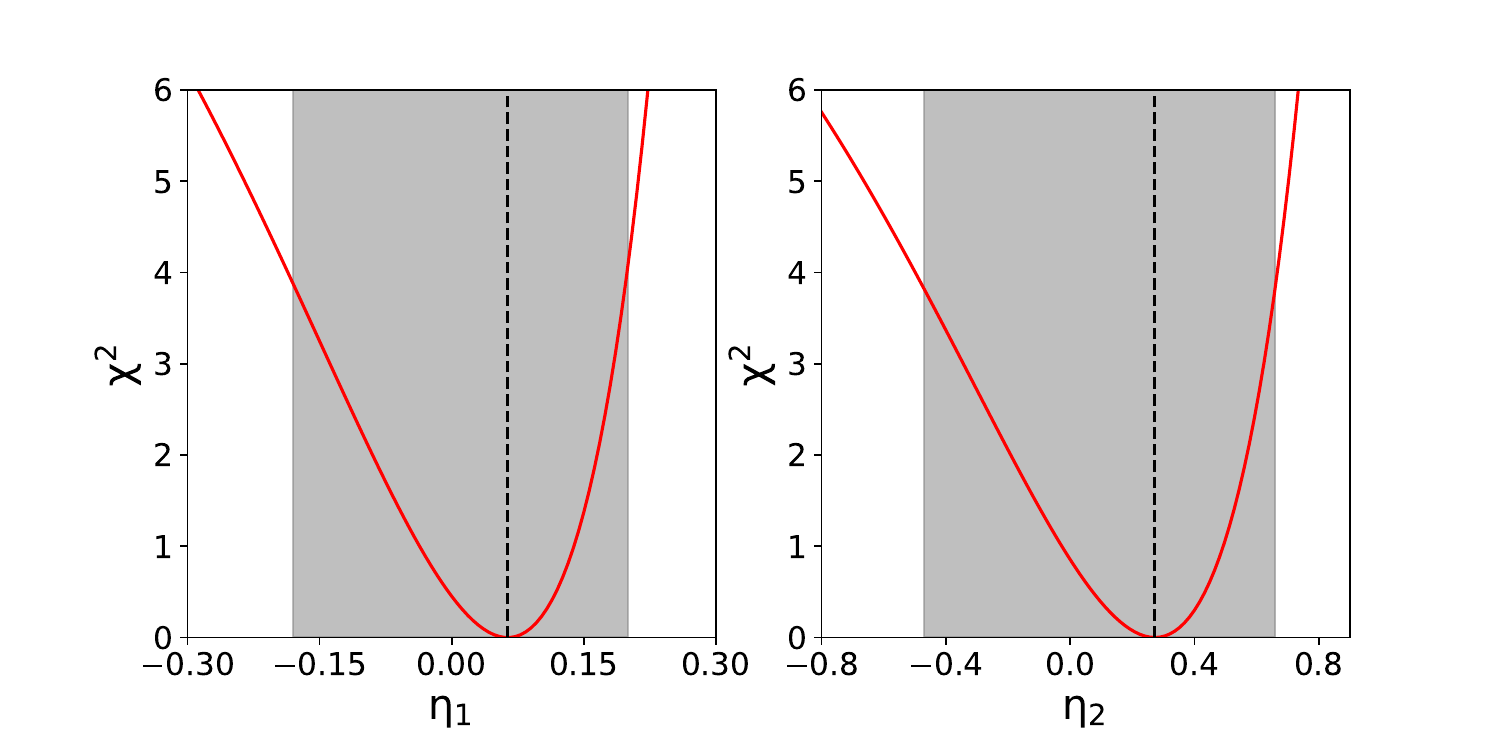}}
%\vskip-0.2in
\caption{$\chi^2$ as a function of the dimensionless LIV parameter $\eta_{n}$ for the linear case (on the left) and the quadratic case (on the right), utilizing the CCF method. The vertical dashed lines indicate the best fits, and the shaded areas represent the 95\% confidence intervals.}
\label{fig:ccf_probdis}
\end{figure}

The $\chi^2$ distribution as a function of unbiased $\eta_n$ is calculated and depicted in Fig.~\ref{fig:ccf_probdis}. {\color{black}The MINUIT package~\cite{1975CoPhC..10..343J}, along with its error estimation processor MINOS~\cite{murtagh1983systems}, is used for fitting and constructing the 95\% confidence levels (by setting $\chi^2$ value 3.84 up the minimum). MINOS accounts for parameter correlations and non-linearities, providing asymmetric error intervals with correct probability coverage for both $\chi^2$ and likelihood fits.} The best-fit values of $\eta_n$ and their uncertainties are translated into limits on the QG energy scale $E_\mathrm{QG}$ at the 95\% confidence level. These limits are detailed in Table~\ref{tab:LIV_results}.

\subsection{Maximum Likelihood Method}

Another approach for inferring $\eta_n$ is the ML method, using Eq.~(\ref{eq:ratelc_func}) as the function to describe the count rate light curve. A binned Poissonian likelihood method is employed for the analysis, with a time binning set to 0.1~s and $N_\mathrm{hit}$ split into 10 segments as illustrated in Fig.~\ref{fig:ten_segments}. The background distribution, as a function of time for each $N_\mathrm{hit}$ segment, is determined using data from the same transit as the GRB, encompassing two sidereal days before and after the burst. The Poisson probability is given by
\begin{equation}
{{\cal P}_{i,j}} = \frac{e^{-(\mu_{{\rm b},i,j}+\mu_{{\rm s},i,j})}\,(\mu_{{\rm b},i,j}+\mu_{{\rm s},i,j})^{N_{{\rm on},i,j}}}{N_{{\rm on},i,j}!},
\label{eq:prob_func}
\end{equation}
where index $i$ represents time bins, and $j$ represents $N_\mathrm{hit}$ segments. The number of signals from the GRB $\mu_{{\rm s},i,j}$ is calculated using Eq.~(\ref{eq:ratelc_func}), the number of background events $\mu_{\mathrm{b},i,j}$ is evaluated from the polynomial fitting of the background as a function of time {\color{black}(see Section~\ref{sec:Appen3} of the Supplemental Material)}, and $N_{\mathrm{on},i,j}$ is the number of observed events. The logarithmic likelihood ratio is defined as
\begin{align}
\varLambda &= -2\ln \frac{{\cal L}_0}{{\cal L}}\nonumber\\
 &=  -2\sum_{i,j}\left[\left(\mu'_{{\rm s},i,j}-\mu_{{\rm s},i,j}\right) + N_{{\rm on},i,j}\ln\frac{\mu_{{\rm b},i,j}+\mu_{{\rm s},i,j}}{\mu_{{\rm b},i,j}+\mu'_{{\rm s},i,j}}\right].
\label{eq:log_lh}
\end{align}
This ratio is utilized for minimization to determine the LIV parameter $\eta_{n}$ and all other light curve parameters. The reference time ($T^\ast$, see the context of Eq.~(\ref{eq:lc_func})) and two parameters for spectral index evolution ($a$ and $b$, see Eq.~(\ref{eq:index_evolution})) are fixed to the values provided in~\cite{2023Sci...380.1390L}. In this equation, ${\cal L} = \prod_{i,j} {\cal P}_{i,j}$ is the likelihood function, and ${\cal L}_0$ is the one for a null hypothesis that no LIV delay exists, where the number of signals corresponds to $\mu'_{{\rm s},i,j}$  through setting $\Delta t_\mathrm{LIV}\left(E,\eta_{n}\right) = 0$ in Eq.~(\ref{eq:ratelc_func}). 

{\color{black}The MINUIT package and MINOS processor is utilized for the likelihood fitting and calculating errors. The best-fit values are $\eta_1 = 0.066$ and $\eta_2 = 0.20$.}

\begin{figure}
%\vskip-0.1in
\centerline{\includegraphics[angle=0,width=0.5\textwidth]{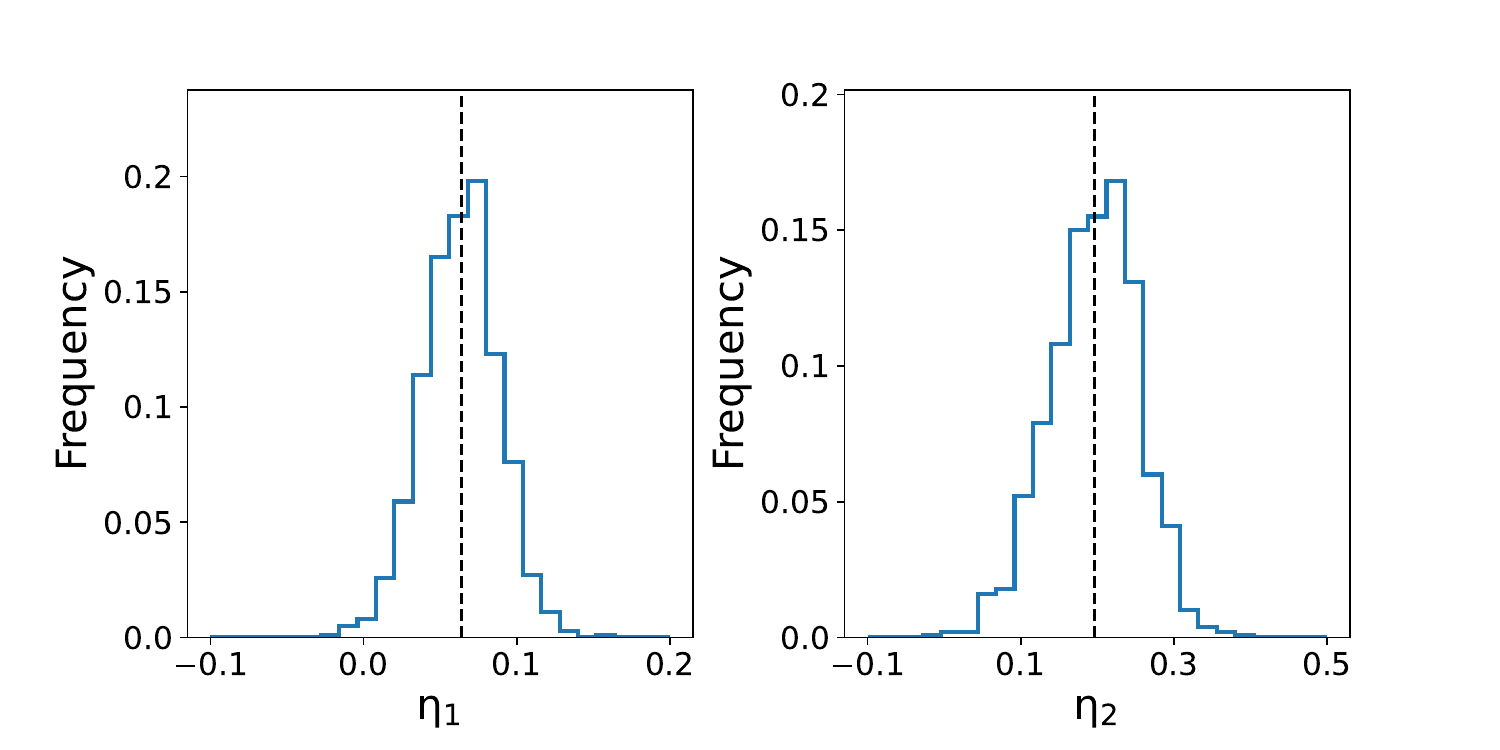}}
%\vskip-0.1in
\caption{Distribution of best-fit values of $\eta_n$ from the shuffled data, presented for both linear (on the left) and quadratic (on the right) cases. The vertical lines indicate the means.}
\label{fig:shuffled_dis}
\end{figure}

\begin{figure}
%\vskip-0.1in
\centerline{\includegraphics[angle=0,width=0.5\textwidth]{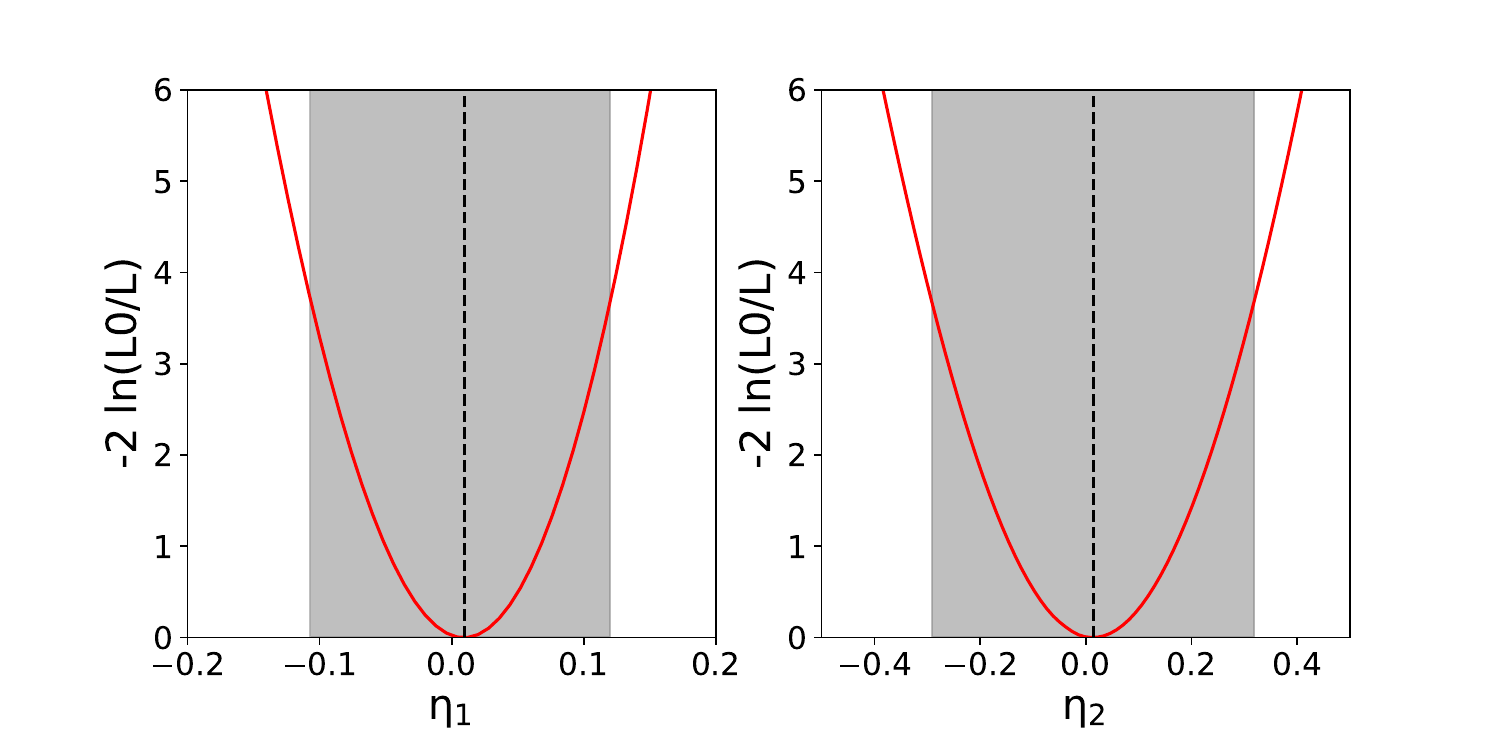}}
%\vskip-0.2in
\caption{Profile likelihood distributions resulting from the analysis with the ML method, depicted for both linear (on the left) and quadratic (on the right) cases. The vertical dashed lines signify the best fits after bias subtraction, and the shaded areas correspond to the 95\% confidence intervals.}
\label{fig:lh_profile}
\end{figure}

\iffalse
\begin{table*}
\renewcommand\arraystretch{1.5}
\tabcolsep=0.4cm
\centering \caption{Values for the best fits (BF) and the 95\% lower (LL) and upper (UL) limits, provided for the dimensionless LIV parameter $\eta_{n}$ using both the CCF and ML methods. Additionally, the 95\% confidence level (CL) lower limits on the quantum gravity (QG) energy scale $E_{\rm QG}$ for the linear ($n=1$) and quadratic ($n=2$) cases are listed.}
\begin{tabular}{lcccccc}
\hline
\hline
Method &  \multicolumn{3}{c}{Cross correlation function} & \multicolumn{3}{c}{Maximum likelihood} \\
\hline
           & $\eta^{\rm LL}$ & $\eta^{\rm BF}$ & $\eta^{\rm UL}$ & $\eta^{\rm LL}$ & $\eta^{\rm BF}$ & $\eta^{\rm UL}$ \\
$\eta_{1}$ & $-0.25$         & $0.05$            & $0.18$            &      $-0.11$           &       $0.01$          &     $0.12$            \\
$\eta_{2}$ & $-0.60$         & $0.25$            & $0.64$            &        $-0.31$         &      $0.02$           &       $0.32$           \\

           & superluminal    &                 & subluminal      & superluminal    &                 & subluminal      \\

$E_{{\rm QG},1}\,[10^{20}\,{\rm GeV}]$  & $0.5$    &                 & $0.7$  &       $1.1$            &               &      $1.0$           \\
$E_{{\rm QG},2}\,[10^{11}\,{\rm GeV}]$ & $5.0$    &                 & $4.8$  &     $7.0$            &                 &      $6.9$           \\
\hline
\end{tabular}
\label{tab:LIV_results}
\end{table*}
\fi

\begin{table*}
\renewcommand\arraystretch{1.5}
\tabcolsep=0.3cm
\centering \caption{Values for the best fits (BF) and the 95\% lower (LL) and upper (UL) limits, provided for the dimensionless LIV parameter $\eta_{n}$ using both the CCF and ML methods. Additionally, the 95\% confidence level (CL) lower limits on the quantum gravity (QG) energy scale $E_{\rm QG}$ for the linear ($n=1$) and quadratic ($n=2$) cases are listed.\footnote{Similar bounds were given in~\citep{2023arXiv230803031P}, which we received while working on this paper.}}
\begin{tabular}{lccccccccc}
\hline
\hline
Method &  \multicolumn{3}{c}{CCF} & \multicolumn{3}{c}{ML (MINOS)} &\multicolumn{2}{c}{ML (Calibrated)}  \\
\hline
           & $\eta^{\rm LL}$ & $\eta^{\rm BF}$ & $\eta^{\rm UL}$ & $\eta^{\rm LL}$ & $\eta^{\rm BF}$ & $\eta^{\rm UL}$ & $\eta^{\rm LL}$ & $\eta^{\rm UL}$\\
$\eta_{1}$ & $-0.18$         & $0.06$            & $0.20$            &      $-0.11$           &       $0.003$          &     $0.12$       & $-0.12$  &  $0.11$    \\
$\eta_{2}$ & $-0.47$         & $0.25$            & $0.66$            &        $-0.31$         &      $0.01$           &       $0.32$       &  $-0.30$  & $0.29$   \\

           & superluminal    &                 & subluminal      & superluminal    &                 & subluminal   & superluminal    &                subluminal   \\

$E_{{\rm QG},1}\,[10^{20}\,{\rm GeV}]$  & $0.7$    &                 & $0.6$  &       $1.1$            &               &      $1.0$   &$1.0$            &                   $1.1$        \\
$E_{{\rm QG},2}\,[10^{11}\,{\rm GeV}]$ & $5.6$    &                 & $4.7$   &     $7.0$            &                 &      $6.9$       & $7.0$  &   $7.2$  \\
\hline
\end{tabular}
\label{tab:LIV_results}
\end{table*}

There could be biases in this approach for $\eta_1$ and $\eta_2$. For instance, Eq.~(\ref{eq:ratelc_func}) does not fit the data well {\color{black}due to the stochastic nature of the afterglow process and various disruptive phenomena involved}. To address this, we analyzed 1000 shuffled data sets by randomly exchanging the time and $N_\mathrm{hit}$ information of events to decouple the correlation between time and energy. To preserve the behavior of the spectral index evolution, we locally shuffled the events within each time bin of 6 seconds and reshuffled them after shifting half of the time bin. The chosen time binning for shuffling is sufficiently large, as the maximum time delay is less than 6 seconds within the energy coverage of the data for $|\eta_1|<0.15$ or $|\eta_2|<0.35$. The means of the 1000 best-fit values of $\eta_n$ from the shuffled data are considered as the biases. As shown in Fig.~\ref{fig:shuffled_dis}, the mean values are $\eta_{\rm 1,bias} =0.063$ and $\eta_{\rm 2,bias} = 0.19$. The profile likelihood curves after subtracting these biases are depicted in Fig.~\ref{fig:lh_profile}.

As no significant LIV time delay from GRB~221009A are detected in this approach, we set limits for $\eta_n$ by constructing 95\% confidence intervals {\color{black}with the assistance of MINOS}. The obtained intervals are subsequently used to compute the limits on the QG energy scale $E_\mathrm{QG}$. The corresponding results are listed in Table~\ref{tab:LIV_results}.

{\color{black} An alternative approach via bootstrapping and shuffling, as developed in~\cite{2013PhRvD..87l2001V,PhysRevLett.125.021301}, is also employed to obtain the so-called ``calibrated'' limits. As shown in the last rows of Table~\ref{tab:LIV_results}, the results appear to be very similar.}

The EBL model could introduce systematic uncertainties in our analysis. We conducted another two rounds of analyses for the same data, considering two extreme cases of the EBL models (corresponding to the upper and lower boundaries of the uncertainty of the EBL model in~\citep{2021MNRAS.507.5144S}). For the linear LIV effect, we observe that the EBL models would enlarge the $\eta_1$ limits by 18\% (12\%) in subluminal (superluminal) scenario. In the quadratic case, the $\eta_2$ limits would be reduced by 6\% (5\%) in subluminal (superluminal) scenario.

\section{Summary}
\label{sec:summary}

LHAASO observed unprecedented large number of VHE photon {\color{black}events} from the brightest GRB 221009A at the earliest epoch, marking the identification of the onset of a TeV GRB afterglow for the first time. These characteristics render this signal a unique opportunity to probe LIV in the photon sector. Utilizing both CCF and ML methods, we searched for LIV-induced lags in the arrival time of the energetic photons. In both methods, compatible limits on the LIV energy scale $E_\mathrm{QG}$ are obtained.

Our limit on the linear modification of the photon dispersion relation is $E_{\mathrm{QG},1}>1.0\times10^{20}\,\mathrm{GeV}$ ($E_{\mathrm{QG},1}>1.1\times10^{20}\,\mathrm{GeV}$), considering a subluminal (superluminal) LIV effect. This is comparable to the most constraining lower limit on $E_{\mathrm{QG},1}$ obtained by the GeV emission of GRB~090510~\cite{2013PhRvD..87l2001V}. In the quadratic case, our result on the energy scale $E_{\mathrm{QG},2}>6.9\times10^{11}\,\mathrm{GeV}$ ($E_{\mathrm{QG},2}>7.0\times10^{11}\,\mathrm{GeV}$) for a subluminal (superluminal) LIV effect represents the best time-of-flight limit, improving previous bounds~\cite{2013PhRvD..87l2001V} by a factor of 5~(7).

Moreover, we emphasize that, thanks to the adoption of the true spectral time lags of bunches of high-energy photons, our constraints on $E_{\mathrm{QG},n}$ could be {\color{black}statistically more robust} compared to previous results of GRB~090510, which were based on the rough time lags of sparse GeV-scale photons.

Future observations of VHE prompt emission instead of afterglow from GRBs would further enhance sensitivity to LIV effects using vacuum dispersion (time-of-flight) tests.

\begin{acknowledgments}
We would like to thank all the members who constructed and operate the LHAASO detectors. This work is supported by the following grants: the National Natural Science Foundation of China (Nos. U1831208, 12005246, 12173039, 12321003, 12373053, and 12375108), the Strategic Priority Research Program of the Chinese Academy of Sciences (grant No. XDB0550400), the Department of Science and Technology of Sichuan Province, China No. 2021YFSY0030, Project for Young Scientists in Basic Research of Chinese Academy of Sciences No. YSBR-061, the Key Research Program of Frontier Sciences of Chinese Academy of Sciences No. ZDBS-LY-7014, the Natural Science Foundation of Jiangsu Province No. BK20221562, and in Thailand by the National Science and Technology Development Agency (NSTDA) and National Research Council of Thailand (NRCT): High-Potential Research Team Grant Program (N42A650868).
\end{acknowledgments}

\bibliographystyle{apsrev4-1}
\bibliography{ms}% Produces the bibliography via BibTeX.

\clearpage

%\onecolumngrid
\appendix

\setcounter{page}{1}
\renewcommand{\thepage}{S\arabic{page}}

\setcounter{figure}{0}
\setcounter{table}{0}
\renewcommand{\figurename}{Figure}
\renewcommand{\thefigure}{S\arabic{figure}}
\renewcommand{\thetable}{S\arabic{table}}
\setcounter{equation}{0}
\renewcommand{\theequation}{S\arabic{equation}}

\section{The Cross-correlation Function and time Lag}
\label{sec:Appen1}

\begin{figure*}
\centering
\begin{subfigure}{0.3\textwidth}\includegraphics[width=\textwidth]{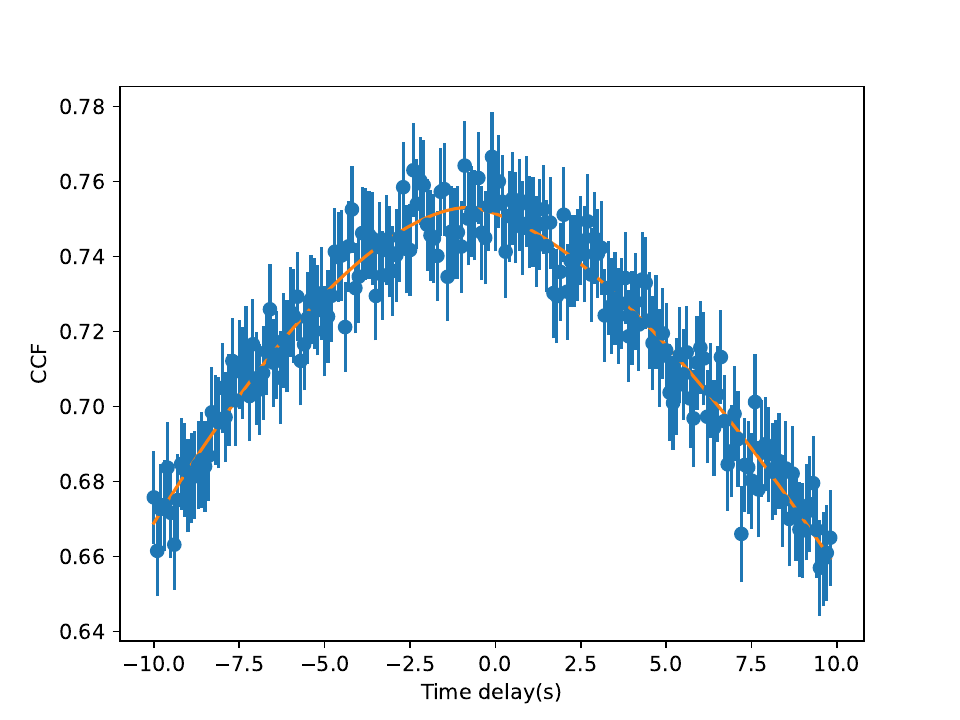}\caption{Seg1 $\bigotimes$ Seg0}\end{subfigure}
\begin{subfigure}{0.3\textwidth}\includegraphics[width=\textwidth]{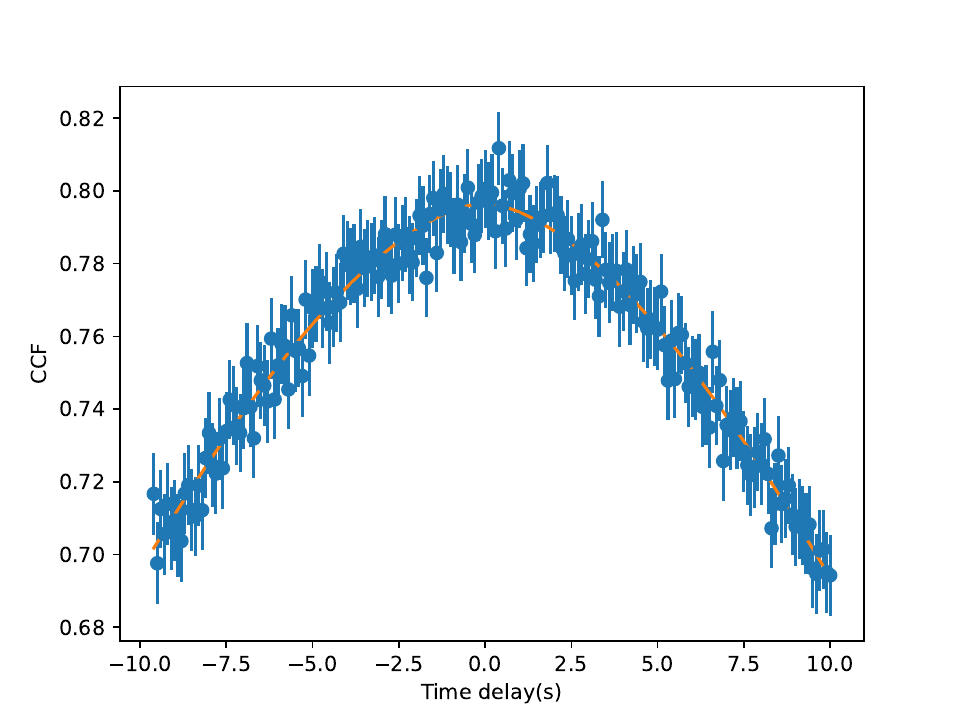}\caption{Seg2 $\bigotimes$ Seg0}\end{subfigure}
\begin{subfigure}{0.3\textwidth}\includegraphics[width=\textwidth]{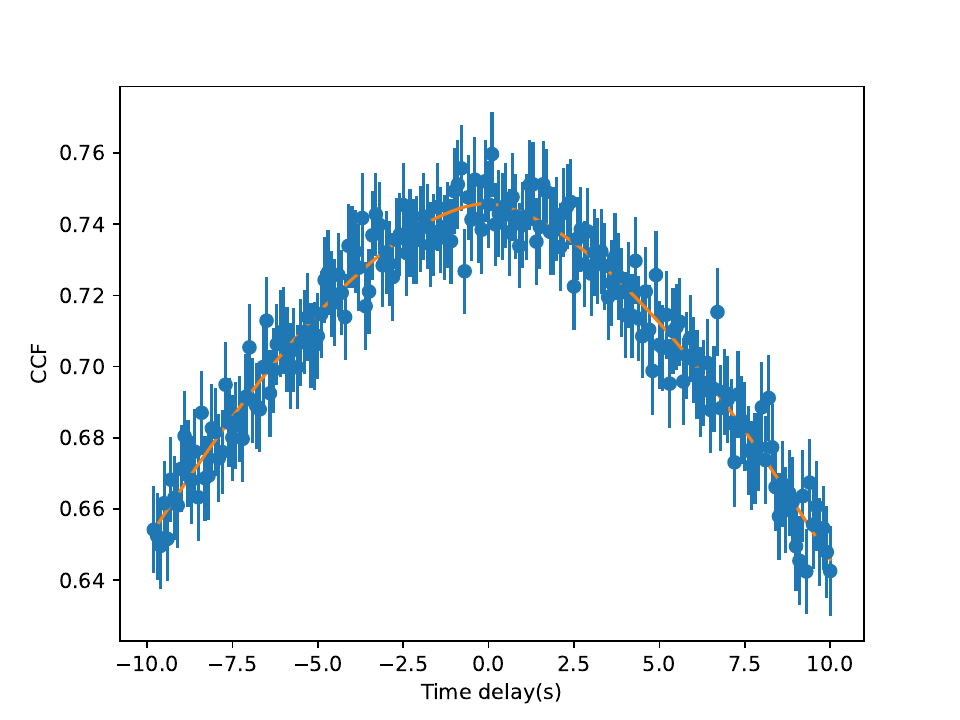}\caption{Seg3 $\bigotimes$ Seg0}\end{subfigure}
\begin{subfigure}{0.3\textwidth}\includegraphics[width=\textwidth]{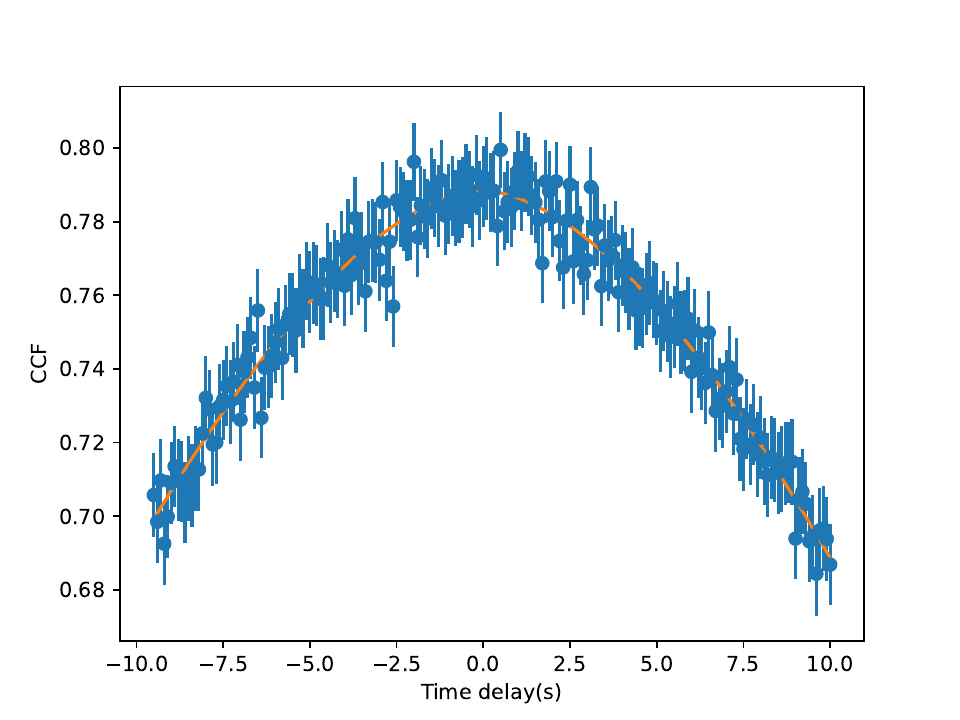}\caption{Seg4 $\bigotimes$ Seg0}\end{subfigure}
\begin{subfigure}{0.3\textwidth}\includegraphics[width=\textwidth]{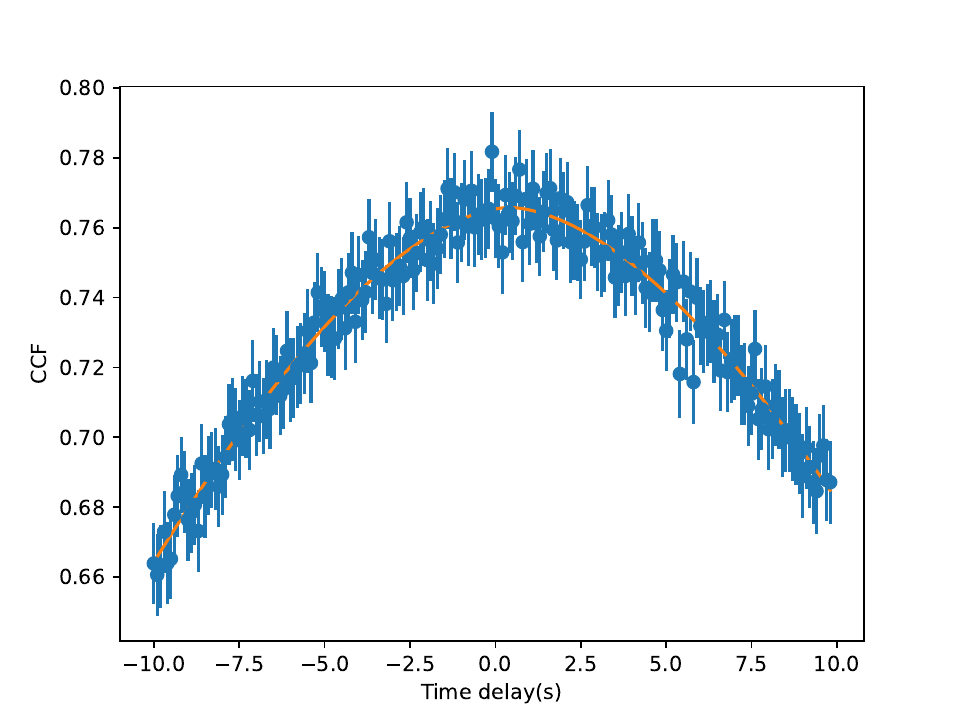}\caption{Seg5 $\bigotimes$ Seg0}\end{subfigure}
\begin{subfigure}{0.3\textwidth}\includegraphics[width=\textwidth]{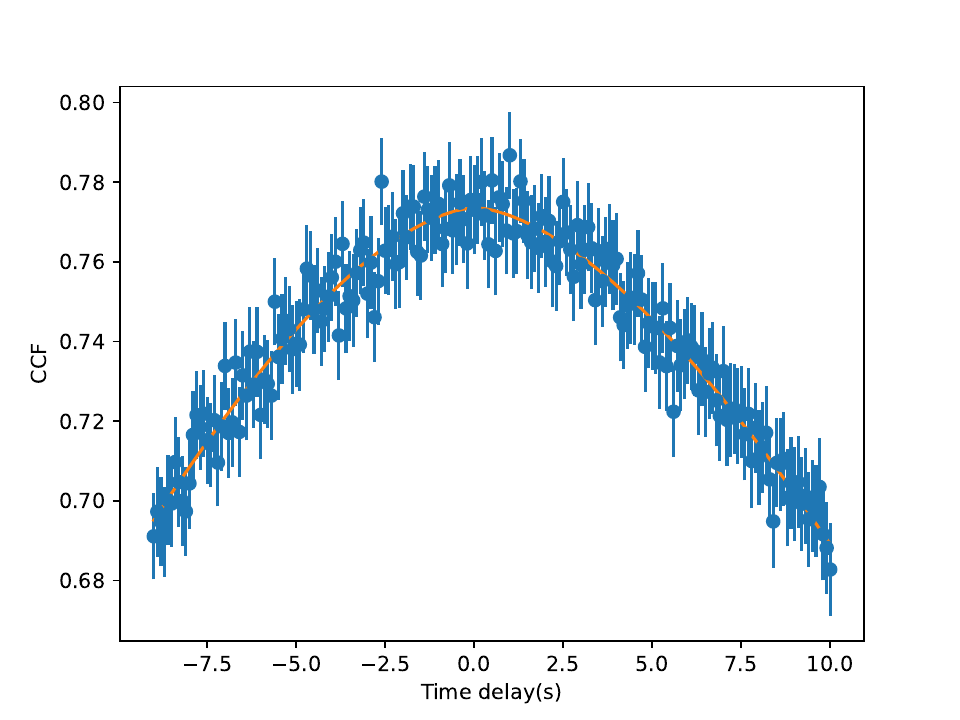}\caption{Seg6 $\bigotimes$ Seg0}\end{subfigure}
\begin{subfigure}{0.3\textwidth}\includegraphics[width=\textwidth]{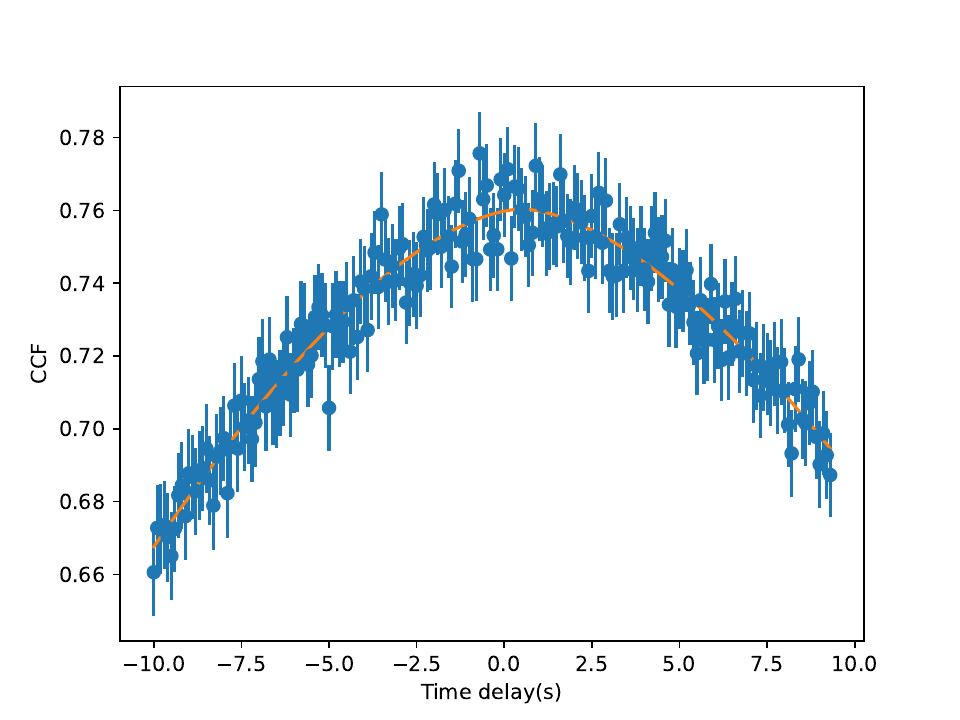}\caption{Seg7 $\bigotimes$ Seg0}\end{subfigure}
\begin{subfigure}{0.3\textwidth}\includegraphics[width=\textwidth]{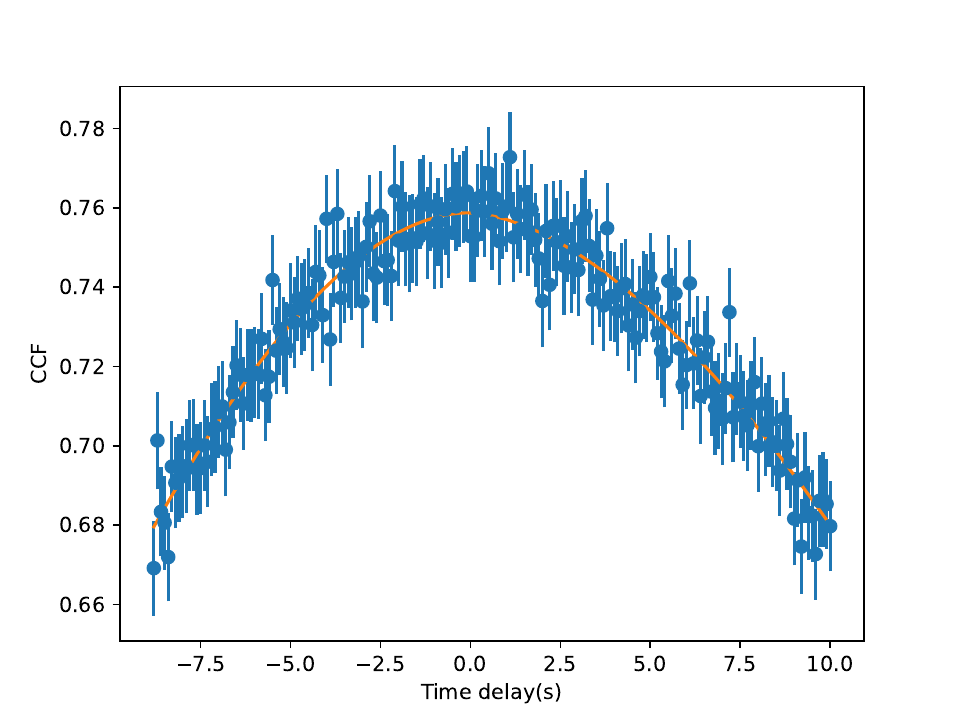}\caption{Seg8 $\bigotimes$ Seg0}\end{subfigure}
\begin{subfigure}{0.3\textwidth}\includegraphics[width=\textwidth]{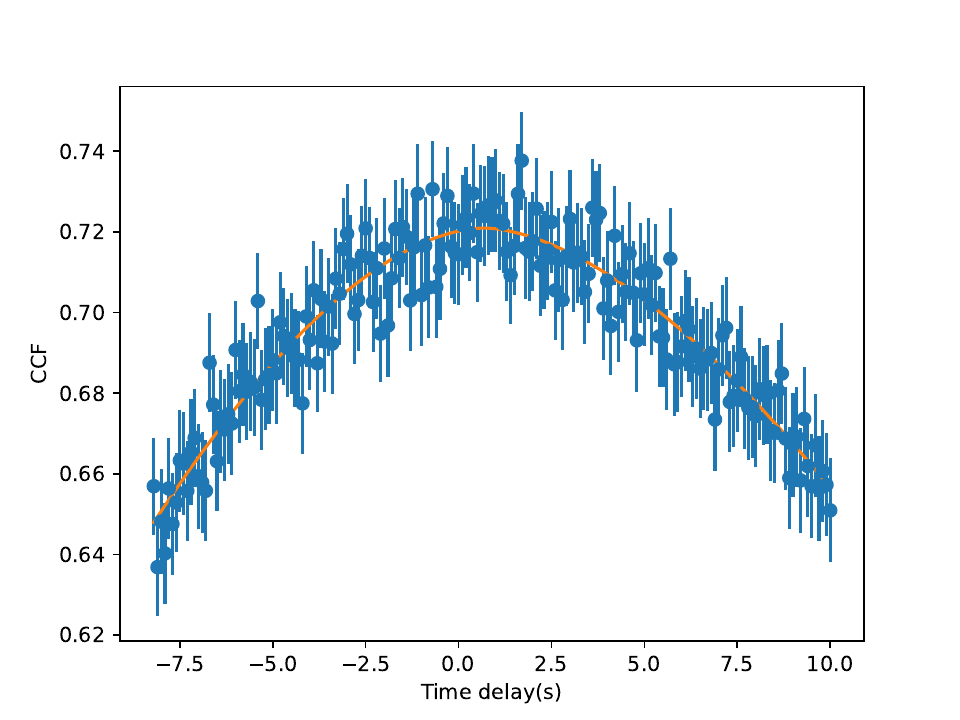}\caption{Seg9 $\bigotimes$ Seg0}\end{subfigure}
\caption{CCF as a function of time delay for the two light curves between the lowest energy band (Seg0) and any of the other nine high energy bands (Seg1--Seg9). The orange line represents the best Gaussian fitting result.}
\label{fig:CCF}
\end{figure*}

For a given pair of light curves, we utilize the cross-correlation function (CCF) to determine the time lag $\Delta t$. 
The CCF with a delay $\Delta t$ is defined as
\begin{align}
F_j(\Delta t) &= \frac{\sum_i R_0(t_i)\,R_j(t_i+\Delta t)}{\sqrt{\sum_i R_0^2(t_i)\,\sum_i R_j^2(t_i)}},
\label{eq:ccf}
\end{align}
where $j = 1, 2, 3, \ldots, 9$ represents the $N_\mathrm{hit}$ segment number except the first one, $i$ is the time bin number, and $R_0(t_i)$ and $R_j(t_i)$ are the rates of segment 0 and $j$ at time bin $t_i$. A discrete analysis is performed with a time step 0.1\,s. 
The corresponding CCF as a function of time delay for the two light curves between the lowest energy band (Seg0) and any of the other nine high energy bands (Seg1--Seg9) are shown in Figure~\ref{fig:CCF}.
We define $\Delta t$ as the time delay corresponding to the global maximum of the CCF. 
By fitting $F_j(\Delta t)$ around the peak with a Gaussian function (see the orange dashed line), a
more precise peak position $\Delta t_j$ can be obtained. From these plots we would say that the distribution around the peak behaves like a Gaussian, and the fitting is quite reasonable, where the obtained peak position reflects the mean behaviour. Some other similar analyses such as
\cite{2010ApJ...711.1073U,2018ApJ...865..153L} also took this kind of treatment.

We employ the {\sl bootstrap} method to estimate the variance of the time lag $\Delta t$, whose square root is taken as the error. The details are as follows. Firstly, we randomly sample these 10 light curves of data using the Monte Carlo (MC) method to obtain 10 mocked distributions, where the size of each MC sample is same as the data. Secondly, using the same CCF analysis method, we calculate the $\Delta t$ for the mocked distributions. Lastly, we repeat above two procedures for 1000 times, and then obtain 1000 $\Delta t$ values, whose variance is finally calculated. This kind of treatment is also adopted by other similar analysis, such as \cite{2010ApJ...711.1073U}.
We assume the systematic uncertainty of $\Delta t$ plays negligible role in the fitting, and only statistical fluctuation takes part. This is the prerequisites of the {\sl bootstrap} method that we applied.

In our CCF analysis method, we extract 10 energy-dependent light curves of GRB 221009A, and the time binning of the light curves used for analysis is 0.1~s.
In order to understand the effects of energy and time binning more fully,
we have ever investigated the cases for three time-bin sizes of 0.1, 0.5, and 1.0~s, and two $N_{\rm hit}$-segment numbers of 5 and 10, and found that the differences in the best-fitted $\eta_n$ (the dimensionless LIV parameters) values of the CCF are less than 2\%, and the differences in the uncertainty of the time lag are less than 5\%. 
Tests show that the choice of energy and time bins for the light curves has negligible impact on the results.

\section{Bias from the Energy Overlapping of Different $N_{\rm hit}$ Segments}
\label{sec:Appen2}
The energy ranges of ten $N_{\rm hit}$ segments are listed in Table~\ref{table:energylimit}. Due to large fluctuation of the cascade processes of low energy particles / photons in the atmosphere, the energy ranges of neighbouring $N_{\rm hit}$ segments are much overlapped. Evaluating the LIV time delay $\Delta t_\mathrm{LIV}(\eta_n)$ based on the median energy may introduce bias, since photon energies exhibit wide dispersion within each $N_\mathrm{hit}$ segment. The procedure of estimating this bias is presented in Section~\ref{sec:method} of the main letter.

\begin{table}
\begin{center}
\caption{\color{black}Energy ranges of 10 $N_{\rm hit}$ segments. The lower and upper limits represent the 16\% and 84\% percentiles of the energy distribution, respectively.}
\label{table:energylimit}
    \begin{tabular}{cccc} 
    \hline
    \hline
    Seg No. & Lower Limit  & Median Energy  & Upper Limit  \\
            &     (TeV)     &     (TeV)      &     (TeV)     \\
    \hline
    0 & 0.153 & 0.354 &0.783\\
    1 & 0.157 & 0.375 &0.828\\
    2 & 0.174 & 0.395 &0.878\\
    3 & 0.189 & 0.419 &0.928\\
    4 & 0.214 & 0.457 &1.012\\
    5 & 0.221 & 0.486 &1.087\\
    6 & 0.256 & 0.556 &1.219\\
    7 & 0.312 & 0.658 &1.435\\
    8 & 0.402 & 0.843 &1.758\\
    9 & 0.732 & 1.601 &3.017\\
    \hline
    \end{tabular}
\end{center}
\end{table}

\section{Polynomial Fitting of the Background}
\label{sec:Appen3}
\begin{figure}
    \centering
    \includegraphics[width=0.8\linewidth]{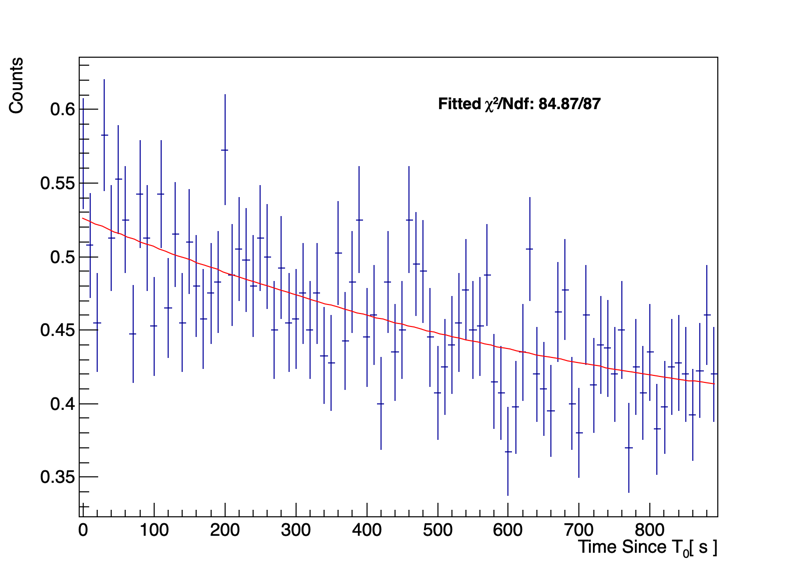}
    \includegraphics[width=0.8\linewidth]{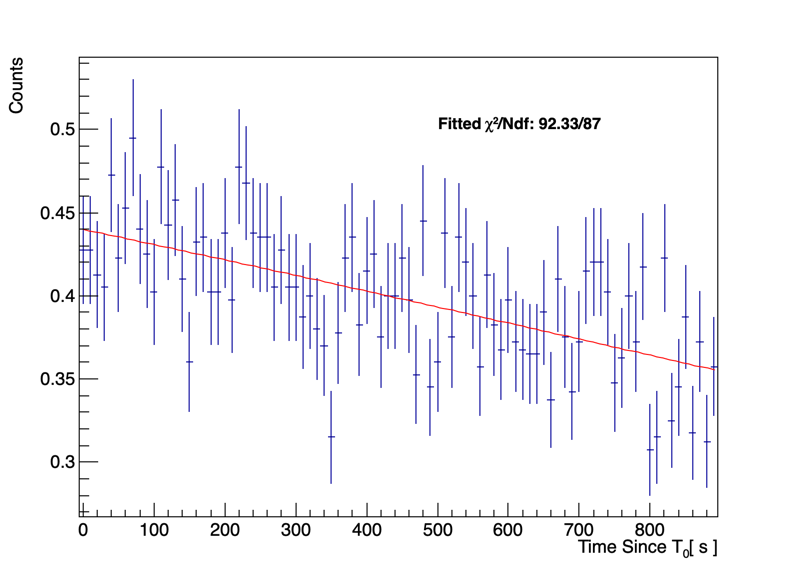}
    \caption{\color{black}Background count rate as a function of time and its polynomial fitting (red line). The upper and lower panel shows the fitting results for $N_{\rm hit}$ segment 0 and 3, respectively.}
    \label{fig:background}
\end{figure}

The polynomial fit to the background has also been adopted in the previous study that was published in~\cite{2023Sci...380.1390L}. Two examples for the background fitting of Segments 0 and 3 are shown in Figure~\ref{fig:background}.

\end{document}